# Research Proposal

# MEMS-EYE: A M/NEMS platform for the investigation of multi-physical and complex nonlinear systems

Proposal by:

Dr. Samer Houri

and

Prof. Gaetan Kerschen

Université de Liege

# 1. DESCRIPTION OF THE RESEARCH PROJECT

## 1.1 Goals of the research

**The ultimate goal of this research proposal is the creation of a micro-opto-mechanical intelligence.**

**The proposal centers on the development and investigation of very large-scale integrated (VLSI) arrays of coupled M/NEMS devices as platforms for the experimental study of nonlinear dynamics of high-dimensional systems. The potential of VLSI M/NEMS arrays to function as advanced sensors will be demonstrated through the novel idea of a MEMS EYE, an electronics-free platform that combines imaging and pattern recognition functionality.**

In the MEMS EYE, the VLSI M/NEMS elements will act as imagers by responding to optical or thermal radiation due to their multi-physical properties, whereas the coupling of these elements will permit them to simultaneously function as a reservoir computing (RC) platform that is capable of image recognition. Thus, the two key aspects of this proposal are (i) the creation and control of coupling between the array elements using various coupling mechanisms, and (ii) the demonstration of the ability of the highly nonlinear and coupled arrays to operate as nonlinear reservoirs for RC.

## 1.2 State of the art

The possibility to produce thousands of similar M/NEMS devices on a millimeter scale substrate represents a singular advantage for VLSI M/NEMS platforms [1-9]. In addition to ease of fabrication, M/NEMS offer a multitude of transduction mechanisms, electrical and optical readout, controllable linear and nonlinear parameters, and various means of coupling.

The very first steps towards the creation of a VLSI M/NEMS platforms have been experimentally investigated using a variety of structural geometries and a variety of materials [1-9]. However, the main thrust behind constructing such arrays has been either to study device variability [5, 7], provide redundancy [4, 8-9], or to increase the signal to noise ratio by averaging over a large number of devices [8- 9]. In these applications, the benefits that can emerge when the elements of the arrays are coupled are not exploited. Coupling can transform the dynamics of a collection of resonators/oscillators and leads to the emergence of global phenomena that cannot be observed otherwise [10-11]. For example, arrays of coupled M/NEMS resonators were used to demonstrate intrinsically localized modes (ILM) [1-2], pattern switching [3], and solitons [12].

Nevertheless, only modest progress has actually been achieved in the study of the dynamics of a large number of coupled M/NEMS devices. **More importantly, no demonstration has been made so far in which the high-dimensional nonlinear dynamics of VLSI arrays contribute to enhance the performance of the M/NEMS devices as sensors or transducers. This is why this project aims to leverage the array-wide dynamics to produce a MEMS EYE, a combined sensing and information processing platform.**

To attain this objective, it is necessary to leverage the three synergistic aspects shown in Fig. 1(a). These are (i) Opto-mechanics [13]: necessary to transduce the image from the optical domain to the mechanical domain. (ii) Coupling: necessary to establish the high-dimensional nonlinear system that will enable RC. And (iii) implementing the RC to allow the coupled arrays to process (visual) information. These three aspects combined will enable the "MEMS EYE".

*Opto-mechanics:*
The optomechanical mechanism used in this work is the opto-thermal effect, i.e., the heating up of the M/NEMS structures due to incident light or radiation. The sensitivity of MEMS devices to incident light is already employed in MEMS bolometers and used for infrared imagers [14-15].

**An important innovation of this work is that the MEMS EYE will exploit opto-thermal interactions not only for the direct detection of light, i.e., imaging, but also as a mechanism for the creation of limit-cycles in M/NEMS resonators [16-17], see Fig. 1(b). As well as using the opto-thermal effect to create novel controllable coupling, Fig. 1(d).**

*Coupling:*
**The creation of coupling between M/NEMS array elements is at the core of this research proposal, and is one of the main experimental difficulties that has delayed progress in the investigation of M/NEMS platforms for reservoirs and high-dimensional dynamical systems.**
In the relevant scientific literature, coupling between array elements has been established either via direct, and uncontrollable, mechanical linkage [1-2, 12] and Fig. 1(c), or electrostatic near-field coupling [3], or near field optical coupling [23], all of which are difficult to control and limited to nearest-neighbor coupling.

This work aims to investigate two coupling mechanisms:
1. Controllable opto-thermal coupling where the opto-thermal effect can be used to couple devices if the light exiting one device can be made to enter another using a spatial light modulator (SLM). This will be the first such demonstration, shown in Fig. 1(d).
2. Internal mode-coupling where many modes of a M/NEMS structure are intrinsically coupled due to nonlinear mechanics. Interestingly, the same mechanics that give rise to this effect in macroscopic structures are valid for microscopic ones. A MEMS network was recently demonstrated using this technique [18].

*Reservoir computing:*
**RC is a new paradigm in the field of artificial intelligence that offers low-power operation and does not require the lengthy training procedure necessitated by typical neural networks.**
These attributes have made RC a prime candidate for IoT "edge-computing", where the low-power information processing happens in-situ at the sensor node [19-22, 24].
To briefly introduce the operating principle of RC, consider a 3D chaotic attractor, say a Lorenz attractor. If inputs to the system are encoded as initial conditions in phase space, then two nearby inputs (similar, but not identical) will quickly diverge in phase space. In other words, the system has separated the inputs. The concept may then be extended to an n-dimensional input, if a corresponding n-dimensional (or more) phase space can be created. An ideal reservoir (for the purpose of RC) would separate inputs while being reproducible, hence states like the edge-of-chaos are considered highly desirable due to their ability to reproducibly separate inputs.
An RC comprises two parts, the reservoir itself and a linear weighted output layer. The reservoir is i) a high dimensional nonlinear dynamical system, with connections that are usually ii) random and iii) sparse. The reservoir is kept unchanged, while the output layer is quickly trained.
RC is surprisingly efficient at the task of separating patterns [24], although despite the oversimplified explanation given above, details regarding the operation of the reservoir should be considered carefully, e.g., dimensions and stability of the reservoir, transient response, etc.

## 1.3    Research Project

Consider the simplest dynamical system, a linear harmonic resonator, now suppose it is possible to tune the frequency of this system, or to change its damping, or to add nonlinear terms of varying orders, or to couple its mechanics to other physical domains; that is the kind of versatile experimental platforms that M/NEMS devices have already been able to provide for the study of multi-physics and nonlinear dynamics and mechanics [25].

Now, suppose we wish to proceed to investigate the dynamics of interacting devices, starting with two coupled MEMS devices and working upwards to high-dimensional dynamical systems containing hundreds of coupled devices, say large numbers of MEMS beams as in Fig.1(c). That is where the M/NEMS community has made little progress in its theoretical and experimental investigations, and that is exactly what this research proposal aims to address.

## Phase 1: GEN1: uncoupled arrays, and the experimental setups

After a brief setting-up period, the scientific activities of the project start with a first generation (GEN1) platform design, in which a very large-scale array of identical, but uncoupled, M/NEMS devices will be fabricated, these will be a single layer M/NEMS devices (one photolithography step) using either SiN/Silicon or SOI wafers and released with HF etching.

The GEN1 platforms will act as a test bed for the development of the basic experimental setups, e.g., transduction (piezo-shaker, photo-thermal) and characterization (Doppler vibrometers, interferometers, and large area camera-based optical readout).

It is known that limit-cycles can be generated in suspended M/NEMS devices by a constant amplitude laser illumination if the proper device parameters are selected [16-17], Fig 1(b). These first-generation platforms will thus also serve to test opto-thermal limit-cycle generation in an array in parallel (a first for a M/NEMS array), thus turning the array of "passive" structures into an array of limit-cycle oscillators. After demonstrating arrays of oscillators, synchronization will be explored via injected signals (again a first for M/NEMS arrays).

While the simplicity of the GEN1 designs means that the fabrication should not be an issue, the necessary readout experimental setups on the other hand will need to be developed. The simultaneous determination of the dynamics of potentially thousands of M/NEMS devices in parallel and at MHz frequency, is a task that will overwhelm any data acquisition system. It is necessary therefore to readout the averaged dynamical response of the VLSI platforms, either the spatial average over all devices, or the time-averaged slow dynamics. Even then, dedicated experimental setups are needed, which include: Photodiode (APD) based fast (array-averaged) readout. Stroboscopic camera based (time-averaged) slow readout, and combinations thereof.

Risk assessment and mitigation: The GEN1 platforms are virtually risk free due to their simplicity. The process of creating limit-cycles and their synchronization on an array level is founded on such well-demonstrated physics that the risks are very limited and issues can quickly be circumvented. A risk exists in the development of the experimental characterization setups as they encompass new features. To spread the risk, a staggered development is envisioned, i.e., start with an APD-based measurement setup (a well-established and risk-free technology) before moving to the camera-based setup (a novel concept).

## *Phase 2: Coupling and the formation of reservoirs*

Having established the basic experimental tools, the work progresses into investigating the effect of coupling on the dynamics, thus moving from a large number of uncoupled low-dimensional nonlinear dynamical systems to a single high-dimensional nonlinear dynamical system. At this point the research activity will bifurcate to follow two different coupling approaches, these are: opto-thermal coupling, and coupling through structural nonlinearities.

The first prong will use the same GEN1 platforms. After establishing during phase 1 that these systems can be turned into arrays of uncoupled limit-cycle oscillators, the necessary coupling to start observing interesting dynamics is then a weak coupling regime. Opto-thermal coupling will thus be used to create the coupling in a controllable manner, by using a reflective spatial light modulator (SLM) to allow light exiting one device to enter another regardless of their geometric proximity, see Fig. 1(d). The second prong is to investigate wide area membrane structures (WAM1) that support a large number of accessible modes, i.e., modes that can be transduced and measured. Due to the inherently present nonlinear mode coupling that exist in M/NEMS devices, these modes will be naturally coupled. The interest of this approach is that it provides an all-to-all type of coupling. The underlying principle behind this approach has only very recently been demonstrated [18].

The two approaches are complementary since the opto-thermal technique is more complex but versatile, whereas the mode-coupling offers simplicity at the expense of a fixed all-to-all coupling.

Risk assessment and mitigation: There is no one critical risk element present in this phase beyond delays, caused either by experimental or funding difficulties. The presence of two prongs give redundancy, and delays can be managed by reorganization of the work packages.

## Phase 3: Implementation of reservoir computing

Once the experimental tools and the underlying VLSI M/NEMS platforms are developed, the next phase is the "training" in which the arrays will be optimized for RC applications.

This phase includes two stages. The first stage follows the standard approach of establishing a random reservoir and training only the output weights ($W_i$ in Fig. 1(e)) to achieve a certain task, say image classification. In this stage, both GEN1 and WAM1 platforms will be used and their performance compared. The second stage explores the possibility to optimize the reservoir itself. To be clear, this is an open question in the RC community. However, the flexibility of the SLM enables the possibility to design the coupling matrix, and thus to compare the performance of reservoirs that have random coupling, exponential coupling, and power-law coupling. This stage is more suited for the opto-thermal coupling approach. In both stages and for both platforms, inputs are supplied via wavelength-multiplexed optically projected patterns, as shown in Fig. 1(e).

An optimized VLSI "MEMS EYE" RC should classify images at a rate of 1000 image/sec, roughly the mechanical settling time of the M/NEMS devices.

Risk management and mitigation: RC and its relation to dynamical systems is well established, the risks are in delays in previous phases. The redundancy in Phase 2 should reduce the risks.

## Phase 4: MEMS EYE demonstrator

The final phase of the 5-year proposal is the demonstration of a "MEMS EYE". The demonstrator will be tested using a mechanical shadow masks in a chopper configuration, and the image classification rate will be used as a benchmark. Afterwards, the developed technology platforms can be used to investigate collective nonlinear dynamical phenomena in M/NEMS arrays.

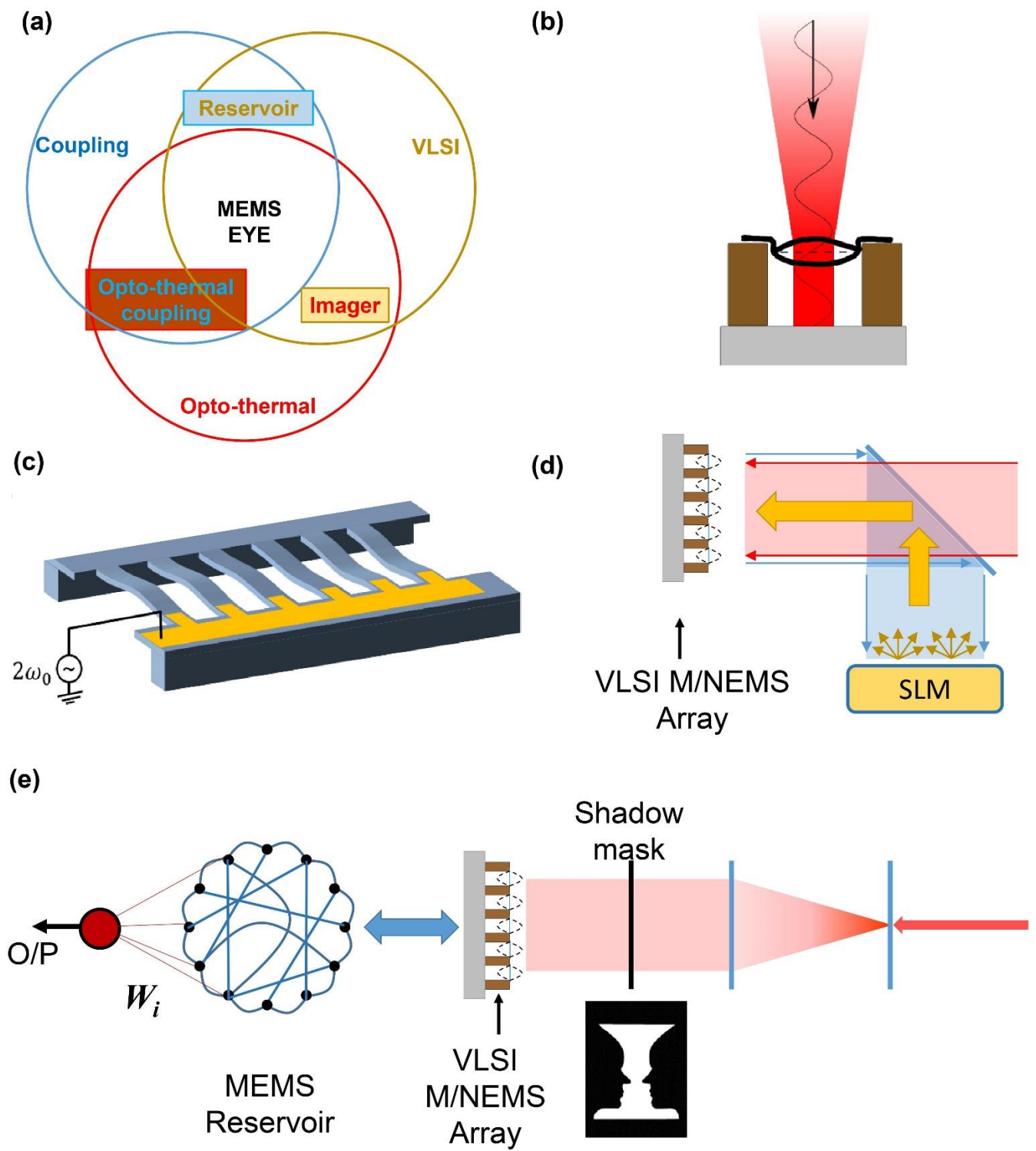

Fig. 1. Schematic representations. (a) Depiction of the multiple research areas involved in a "MEMS EYE". (b) Opto-thermal induced limit-cycles. Incident laser

illumination is reflected from the substrate forming a partial standing wave with position-dependent heating effect. The structure's thermal and mechanical time constants are such that the heating induces bending and the cooling is out of phase with the mechanical motion, thus the initial position becomes unstable and a limit-cycle is generated. (c) An array of mechanically coupled M/NEMS beam structures, the coupling is due to undercut near the anchoring of the beams. From [12]. (d) Photo-thermal coupling. Light that is reflected off the M/NEMS array is guided onto a SLM through a series of optics (not shown), gets amplitude-phase modulated by the SLM and reflected onto the VLSI array. (e) The operation principle of the "MEMS EYE". Incoming light projected through a shadow mask (input pattern) opto-thermally interacts with the M/NEMS structures. The M/NEMS devices are coupled and form a reservoir. The reservoir responds as a whole to the input pattern, its overall dynamics are measured and an output is produced that is dependent on the input pattern.